\documentclass[prd,preprint,nofootinbib,a4paper]{revtex4}

\usepackage{amsmath,amssymb}
\usepackage{graphicx}
\usepackage{bbold}

\newcommand{\md}{\mathrm{d}}
\newcommand{\me}{\mathrm{e}}
\newcommand{\mcl}{\mathcal{L}}
\newcommand{\mcm}{\mathcal{M}}

\newcommand{\mcs}{\mathcal{S}}
\newcommand{\Guone}{\ensuremath{\mathrm{U}(1)}}

\newcommand{\Gsutwo}{\ensuremath{\mathrm{SU}(2)}}
\newcommand{\Gsuthree}{\ensuremath{\mathrm{SU}(3)}}
\newcommand{\parpar}[2]{\frac{\partial#1}{\partial#2}}
\newcommand{\symop}{\boldsymbol\alpha}

\DeclareMathOperator{\floor}{floor}
\DeclareMathOperator{\Tr}{Tr}

\DeclareMathOperator{\pr}{pr}

\newcommand{\half}{\frac{1}{2}}

\newenvironment{caselist}{\begin{list}{}
{\setlength{\labelwidth}{2.7cm}
\setlength{\labelsep}{0.4cm}
\setlength{\itemindent}{0pt}
\setlength{\leftmargin}{3.2cm}
\setlength{\rightmargin}{1cm}
\setlength{\parsep}{0.5ex plus0.2ex minus0.1ex}
\setlength{\itemsep}{0.3ex plus0.2ex}}}
{\end{list}}

\begin{document}


\title{A systematic approach to model building}

\author{Damien P. George}
\email{dpgeorge@nikhef.nl}
\affiliation{Nikhef Theory Group, Science Park 105, 1098 XG Amsterdam, The Netherlands}

\date{\today}
\preprint{NIKHEF/2011-015}

\begin{abstract}
We outline a new, systematic way of constructing and analysing
field theories, where all possible continuous symmetries of a
given model are derived using the method of Lie point symmetries.
If the model has free parameters, and relationships amongst any of
these parameters yields an enhanced symmetry, then all such
relationships are found, along with the resulting symmetry group.
We discuss how the method can be applied to the standard model and
beyond, to direct the search for a more predictive field theory.
The method handles compact and non-compact continuous groups,
spontaneously broken symmetries, and is also applicable to general
relativity.
\end{abstract}

\maketitle


\section{Introduction}

If one wanted to compute all the possible symmetries of a
Lagrangian, one could naively write all coordinates and fields as
arbitrary functions of every coordinate and field, demand that
the resulting Lagrangian was the same as the original, and then
solve for the arbitrary functions.  As long as one can obtain all
the solutions, this method would give an \emph{exhaustive} list of
the symmetries of the model.  Furthermore, if there are free
parameters in the model that, when given specific values or
relationships to other parameters, yield a different set of
symmetries, then the method would necessarily cover these cases as
well.

The obvious problem with such a method is that one would obtain
equations which are just as, if not more, difficult to solve than
the entire system itself.  But the general idea is a powerful one
and we would like to implement it in at least some reduced form.
In fact, if we are content with finding only continuous symmetries
then, by the virtue that a continuous group is classified by its
local Lie algebra, the arbitrary functions mentioned above need
only be infinitesimal and the resulting set of equations that need
to be solved are linear partial differential equations (PDEs).

Such a technique exists and was originally worked out by Lie in order to
assist in solving differential equations.  If one knows a certain
solution of a set of PDEs, then knowing the symmetries allows one
to easily construct other solutions.  This was the motivation
behind Lie's development of the theory.  The method is now known by
various names depending on how general a symmetry one is looking
for.  Here we shall be concerned with the Lie point symmetry
(LPS), which is a continuous symmetry that can depend on
coordinates and fields, but not on derivatives of the fields.  The
LPS method involves finding and solving a set of partial
differential equations known as the determining equations.  These
``determine'' the allowed symmetries, and, when solved in a
systematic way, yield all interesting relationships between free
parameters of the original system.

The basic text book is by Olver~\cite{Olver:1986aa} and discusses
in depth most of the theory related to finding symmetries.  A
useful algorithm for solving a large set of PDEs is described in
detail by Reid~\cite{Reid:1990aa}, and further improvements on
this are given in Refs.~\cite{Reid:1991aa, Reid:1992aa}.  The idea
of using the LPS method to find all the symmetries of a field
theory has previously been applied to the case of scalar QED and
Weyl QED~\cite{Hereman:1993aa}, to Einstein's
equations~\cite{Marchildon:1995ma}, and to Yang-Mills in
4d~\cite{Marchildon:1997xc}.  The extension to supersymmetry has
also been developed and utilised~\cite{Grundland:2008ak}.  
A comprehensive survey of this field is given
in~\cite{Hereman:1996aa}, with an emphasis on available
computer programs to automatically carry out the LPS method.
These past studies have overlooked the utility of the method for
particle physics and model building, in particular the ability
to find parameter relationships that yield a larger symmetry
group.

In this paper we shall show in detail how the LPS method works and
discuss its application to field theories.  Using these ideas one
has a new approach to model building, where symmetries and
parameter relationships are systematically \emph{derived}, not
input from the start.  We begin in Section~\ref{sec:lps} with a
description of the LPS method, followed by a simple example in
Section~\ref{sec:2s}.  In Section~\ref{sec:nsc} we derive the
allowed symmetries of a general theory with $N$ interacting
scalars, and specialise to cases with low $N$.
Section~\ref{sec:act-vs-eom} discusses, with the aid of a worked
example, the difference between the symmetries of the action and
those of the equations of motion.  We make some remarks on the
utility of the LPS method in Section~\ref{sec:remarks}, including
its ability to handle spontaneously broken symmetries.  A way to
automate the method is outlined in Section~\ref{sec:auto}, along
with a plan for the construction of a catalogue of field theories.
The ultimate aim is to be able to apply the LPS method to the
standard model, and beyond, to give a more directed search to
new theories of particle physics, a topic which is discussed in
Section~\ref{sec:sm}.  We conclude the paper in
Section~\ref{sec:concl}.


\section{The Lie point symmetry method}
\label{sec:lps}

For a given system, the Lie point symmetry method consists of
finding the associated determining equations, whose solutions
describe infinitesimal symmetries, and then solving these
equations.  The term ``point'' means that the finite
transformations of the coordinates and fields depend only on the
coordinates and fields themselves, and not the derivatives of the
fields.  According to Lie, one does not have to look at the full
finite transformation; it is enough to study their infinitesimal
behaviour.  The method proceeds as follows.
\begin{enumerate}

\item Derive the determining equations of the system.

Given an action, or equations of motion, with coordinates
$x^\mu$ and fields $\phi_i$, one makes a general infinitesimal
variation of the coordinates, $\delta x^\mu=\eta^\mu$, and
fields, $\delta \phi_i=\chi_i$, and obtains equations~--- the
so-called determining equations~--- which are linear partial
differential equations for $\eta^\mu$ and $\chi_i$.  The solutions
of these determining equations describe the set of symmetries of
the original system.

\item Solve the determining equations, or at least reduce them to
a standard form.

This is an involved step.  For simple systems the determining
equations are only weakly coupled and can be completely solved,
or reduced to algebraic equations, with little effort.  This is
not the case for more complicated systems.  However, due to the fact
that the determining equations are linear, there is a well-defined
algorithm which completes in finite time and brings the determining
equations to a standard form, where they are in involution.  The
algorithm is in essence Gaussian elimination: an ordering on
$\eta^\mu$ and $\chi_i$ is defined, the terms of the equations are
sorted using this ordering, each equation is treated as a row,
and the system is reduced to ``diagonal'' form.

A critical part of the reduction is the ``column elimination''
step.  The coefficient of the leading term used in the elimination
is in general a function of the free parameters of the original
model.  If the values of the parameters have specific values such
that this coefficient is zero, this particular elimination cannot
proceed.  The algorithm must then branch, with one branch of the
solution corresponding to the coefficient being zero, and the
other branch, being non-zero.  The algorithm then continues with
each branch independently, possibly spawning additional branches
as the reduction proceeds.  In general, the resulting solution of
each branch, and hence the symmetries, are different.

This branching is a general feature of the LPS method.  Given a
model, for each set of relationships among its parameters that
yield a different symmetry group the LPS method will produce a
branch associated with this set, and the parameter relationship
will be specified.

\item Compute the rank of the symmetry set(s).

For each branch we compute the amount of symmetry~--- the
rank~--- by counting the number of integration constants (the
initial data) in the reduced set of determining equations.  If the
equations of the branch are solved then one has already found
these integration constants.  If the branch is not solved, but is in standard
reduced form, one can unambiguously compute the amount of initial
data needed to fully specify a unique solution to the set of
differential equations.  Either way, the amount of initial data
gives the rank of the symmetry corresponding to the solutions of
that branch.

The notation for the rank is a tuple of integers
$R=(N_\text{const},N_{f_1},N_{f_2}\ldots)$.  Here,
$N_\text{const}$ is the number of independent constants, $N_{f_1}$
the number of independent functions of one variable, $N_{f_2}$ the
number of functions of two variables, and so on.  Trailing zeros
in the tuple will generally be suppressed.  Each constant and
function parameterises an independent symmetry, so the rank gives
a compact and precise indication of the amount of symmetry in a
given system.

\item Compute the action of the symmetries.

This last step is not always possible, and only necessary if one
wants to know how a certain symmetry acts on the coordinates and
fields.  One must solve explicitly for the forms of $\eta^\mu$ and
$\chi_i$, if they have not already been obtained, and then
compute the action of the symmetry on the coordinates and fields.
One uses the usual techniques for solving differential equations
to do this.

\end{enumerate}

As we shall demonstrate, the LPS method is a very general and
powerful tool, and allows one to build and analyse field theories
in a completely systematic way.  We emphasise that the method
\begin{itemize}
\item is an exhaustive search of continuous symmetries,
\item yields all interesting relationships between parameters, and
\item is guaranteed to terminate (up to the end of step 3) in 
finite time, determined by the number of coordinates and number of
fields.
\end{itemize}
Of course, it has some drawbacks.  There is no guarantee that one
can solve for the actions of the symmetries (step 4), although it
is possible to solve for the structure constants of the
group~\cite{Reid:1992aa}.  Apart from this, the biggest disadvantage
of the LPS method is that the number and complexity of determining
equations increases rapidly (but still polynomially) with the
number of fields $\phi$.  The number of equations also increases
drastically with the number of coordinates $x$, but this is not
such a problem if one sticks with 4d
theories.\footnote{If one can show, as we do for a specific set of
theories in Section~\ref{sec:nsc}, that the symmetries of the
coordinate and field sectors separate, then having a large number
of dimensions is no problem.}
The growth with number of fields is our major concern, and it
seems that the only way forward is to automate the above 3 (or 4)
steps.  Actually, these steps lend themselves quite nicely to
automation, as we shall discuss in Section~\ref{sec:auto}.

We now describe in more detail step 1: how to derive the set of
determining equations for a given system.  One can start from the
action, or from the equations of motion.


\subsection{The action approach}

Given an action, one makes a general variation by adding
infinitesimals to all the coordinates and fields.  Demanding that
the result is equivalent to the original action gives the master
determining equation, from which one obtains the individual set of
determining equations.   We shall derive the form of the master
equation for a general Lagrangian that depends only on
coordinates, fields and first derivatives of fields.  For more
detail see ~\cite{Olver:1986aa, Torres:2004aa}.

Consider then a general action of real fields $\phi_i(x)$ (which
could be the components of a field with arbitrary spin properties)
in arbitrary dimensions:
\begin{equation}
  \mcs = \int \md^nx \; \mcl\left[x^\mu, \phi_i(x), \partial_\mu \phi_i(x)\right] \:.
\label{eq:get-act}
\end{equation}
The infinitesimal point transformation is
\begin{equation}
  x^\mu \to \bar{x}^\mu = x^\mu + \eta^\mu(x,\phi) \:,\qquad
  \phi_i \to \bar{\phi}_i = \phi_i + \chi_i(x,\phi) \:.
\end{equation}
Under this transformation one can show that the action
transforms to
\begin{equation}
\mcs \to \int \md^nx \left[
  \mcl
  + \mcl \frac{\md\eta^\mu}{\md x^\mu}
  + \parpar{\mcl}{x^\mu} \eta^\mu
  + \parpar{\mcl}{\phi_i} \chi_i
  + \parpar{\mcl}{(\partial_\mu\phi_i)} \left(
    \frac{\md\chi_i}{\md x^\mu}
    - \parpar{\phi_i}{x^\nu} \frac{\md\eta^\nu}{\md x^\mu}
  \right)
\right] \:.
\label{eq:acttrans}
\end{equation}
Sum over repeated $\mu,\nu$ and $i$ indices is understood.  The
total derivative is
\begin{equation}
\frac{\md}{\md x^\mu}
  \equiv \parpar{}{x^\mu}
  + \parpar{\phi_i}{x^\mu} \parpar{}{\phi_i} \:.
\end{equation}
In deriving this equation we have used the Jacobian transformation
matrix for the coordinates,
$J=\partial\bar{x}/\partial x=\mathbb{1}+\partial\eta/\partial x$,
which transforms $x$ to $\bar{x}$.  Some useful properties are
$J^{-1}=\mathbb{1}-\partial\eta/\partial x$ and
$\det(J)=1+\Tr(\partial\eta/\partial x)$.\footnote{
Note that if we take equation~\eqref{eq:acttrans}, set $\eta=0$,
and do integration by parts on one of the remaining terms, then we
can obtain the usual Euler-Lagrange equations for the $\phi_i$.}

The infinitesimals $\eta$ and $\chi$ do not necessarily vanish on
the boundary (for example, a time translation symmetry has
$\eta^t$ a constant) so we cannot do integration by parts
on~\eqref{eq:acttrans}.  This equation is thus in its simplest
form, and the condition for the transformation to leave the action
invariant is\footnote{One could equate this to a total coordinate
derivative to give more freedom, and possibly additional
symmetries.  We do not consider such a case here.}
\begin{equation}
  \mcl \frac{\md\eta^\mu}{\md x^\mu}
  + \parpar{\mcl}{x^\mu} \eta^\mu
  + \parpar{\mcl}{\phi_i} \chi_i
  + \parpar{\mcl}{(\partial_\mu\phi_i)} \left(
    \frac{\md\chi_i}{\md x^\mu}
    - \parpar{\phi_i}{x^\nu} \frac{\md\eta^\nu}{\md x^\mu}
  \right) = 0 \:.
\label{eq:actde}
\end{equation}
This is a key result.  It is the master determining equation.
For a given Lagrangian density one computes the above expression
explicitly, treats all the \emph{derivatives} of the fields
$\phi_i$ as independent variables, and then equates the
independent coefficients to zero.  The resulting equations are the
determining equations.  Note that the result can be extended in a
straightforward way to the case where the Lagrangian density
depends on higher order derivatives of the fields.

Once $\eta^\mu$ and $\chi_i$ have been solved for, one can obtain
the finite action of the symmetry group by solving the coupled
differential equations
\begin{equation}
  \frac{\md \bar{x}^\mu}{\md \epsilon} = \eta^\mu(\bar{x},\bar\phi) \:,\qquad
  \frac{\md \bar\phi_i}{\md \epsilon} = \chi_i(\bar{x},\bar\phi) \:,
\label{eq:eta-chi-sym}
\end{equation}
where $\epsilon$ is the continuous group parameter.  The initial
conditions for this set of equations are $\bar{x}^\mu(0)=x^\mu$ and
$\bar\phi_i(0)=\phi_i$.

In summary, the idea is the following.
Given an action, one varies it by adding
infinitesimals to the variables and demanding that the resulting
action is the same as the original.  In the usual case, one lets
the \emph{infinitesimals be arbitrary and solves for the field
configuration}.  This yields the Euler-Lagrange equations which
are classical solutions that extremise the action.  The situation
here is the reverse: in equation~\eqref{eq:actde} we let the
\emph{field derivatives be arbitrary and solve for the
infinitesimals}.  This gives the determining equations, which are
a set of coupled \emph{linear and first order} PDEs.\footnote{The
determining equations coming from the Euler-Lagrange equations are
in general second order because equations of motion have second
derivatives of the fields in them.  Here the determining equations
are first order because $\mcl$ (by assumption) has only first
derivatives in it.}
In some sense we have, in this way, obtained a linear version of
the theory, whose solution gives all the symmetries of the action.
And, as we have pointed out before, when finding the symmetries
all the interesting relationships between parameters of the model
become apparent in a systematic way.


\subsection{The equations of motion approach}
\label{sec:lps-eom}

It is also possible to obtain the determining equations from the
equations of motion of the system, the Euler-Lagrange equations.
This approach requires more effort than the action approach since
an action with only first-order derivatives of the fields will in
general have second-order equations of motion.   Nevertheless, it
is worth discussing this alternative technique since it is suited
to finding continuous symmetries of any set of PDEs, not just
those of a field theory.

The system under consideration can be a function of both
independent, $x^\mu$, and dependent, $\phi_i(x)$, variables,
and can include derivatives of the $\phi_i$.  Denote the system by
\begin{equation}
  \Delta_j(x^\mu,\phi_i,\partial\phi_i)=0 \:,
\end{equation}
where $j$ indexes each equation of the system and
$\partial\phi_i$ can be a derivative of arbitrary order.  Given
this system of PDEs, the determining equations are obtained as
follows.
\begin{enumerate}
\item Construct the prolonged symmetry operator $\pr^{(k)}\symop$.

Point symmetries of the system described by $\Delta$ take
the form $x^\mu\to\bar{x}^\mu=X^\mu(x,\phi)$,
$\phi_i\to\bar{\phi}_i=\Phi_i(x,\phi)$, with
$\Delta_j(X_i,\Phi_i,\partial\Phi_i)=0$.
We construct the differential operator $\symop$ which, when
applied to an object, realises the infinitesimal point symmetry
transformation:
\begin{equation}
\symop =
  \eta^\mu \parpar{}{x^\mu}
  + \chi_i \parpar{}{\phi_i} \:,
\end{equation}
with implicit sum over the indices.  Because the system $\Delta$
can depend on derivatives of $\phi$, the operator $\symop$ must be
prolonged to $\pr^{(k)}\symop$ so that it acts correctly on this
extended space of functions.  $k$ here is the highest order
derivative of the system.  The formula for $\pr^{(k)}$ is
complicated and will not be given here; see~\cite{Olver:1986aa}
or~\cite{Hereman:1996aa}.  It essentially extends $\symop$ to
include all possible combinations of derivatives of $\phi$, to
order $k$.

\item Apply $\pr^{(k)}\symop$ to the system.

We demand the following holds:
\begin{equation}
(\pr^{(k)}\symop\cdot\Delta)\rvert_{\Delta=0}=0 \:.
\label{eq:prdelta}
\end{equation}
This means that the action of the infinitesimal symmetry
generator leaves the system unchanged, when evaluated on a
solution.

\item Obtain the determining equations by equating all independent
coefficients to zero.

The independent variables in equation~\eqref{eq:prdelta} are
derivatives $\partial\phi$ (since $\eta$ and $\chi$ only depend on
$x$ and $\phi$).  Thus~\eqref{eq:prdelta} holds in general only
when all the coefficients of these derivatives vanish.  Extracting
these coefficients and setting them to zero yields the determining
equations.  These equations are a set of linear PDEs in the
variables $\eta$ and $\chi$, which both depend on $x$ and $\phi$.

\end{enumerate}

A more detailed description of these steps can be found in,
for example, \cite{Olver:1986aa} or \cite{Hereman:1996aa}.
We shall give an example using the equations of motion approach in
Section~\ref{sec:act-vs-eom}, and also discuss its differences to
the action approach.


\section{An example: two scalars and a $\Guone$ symmetry}
\label{sec:2s}

The best way to understand the LPS method is to work through an
example.  Let us do that by finding the symmetries of two massive
scalars (spin-0 fields).  We shall ignore the coordinate sector to
keep things simple.  This still allows us to see the workings of
the method and an example of branching for different values of
parameters, in this case the masses.

The Lagrangian density in an arbitrary number of dimensions is
\begin{equation}
  \mcl =
    \half \partial^\mu\phi_1 \partial_\mu \phi_1
    + \half \partial^\mu\phi_2 \partial_\mu \phi_2
    - \half m_1^2 \phi_1^2
    - \half m_2^2 \phi_2^2 \:.
\end{equation}
There are two independent field variables and we must solve for
their variations, $\chi_1(\phi_1,\phi_2)$ and
$\chi_2(\phi_1,\phi_2)$.
The master determining equation is found by feeding the above
Lagrangian density into equation~\eqref{eq:actde}.  One obtains
\begin{equation}
  -m_1^2\phi_1\chi_1 - m_2^2\phi_2\chi_2
  + \partial^\mu\phi_1\partial_\mu\phi_1\parpar{\chi_1}{\phi_1}
  + \partial^\mu\phi_1\partial_\mu\phi_2\parpar{\chi_1}{\phi_2}
  + \partial^\mu\phi_2\partial_\mu\phi_1\parpar{\chi_2}{\phi_1}
  + \partial^\mu\phi_2\partial_\mu\phi_2\parpar{\chi_2}{\phi_2} = 0 \:.
\end{equation}
Since the $\chi$'s do not depend on derivatives of the fields we
must have the coefficients of all independent derivative factors
vanish.  This gives the four determining equations for our system:
\begin{equation}
\begin{aligned}
  & \parpar{\chi_1}{\phi_1} = 0 \:,\qquad
  \parpar{\chi_1}{\phi_2} + \parpar{\chi_2}{\phi_1} = 0 \:,\qquad
  \parpar{\chi_2}{\phi_2} = 0 \:,\\
  & -m_1^2\phi_1\chi_1 - m_2^2\phi_2\chi_2 = 0 \:.
\end{aligned}
\end{equation}
These equations are simple enough that we can solve them directly.
The first three equations give the general solution
\begin{equation}
  \chi_1(\phi_2) = \alpha_1 + \beta \phi_2 \:,\qquad
  \chi_2(\phi_1) = \alpha_2 - \beta \phi_1 \:,
\label{eq:2s-chi-soln}
\end{equation}
with $\alpha_{1,2}$ and $\beta$ constants.  As we shall see, each
of these correspond to a specific symmetry.  The maximum rank of
the system is $R=(3)$ because we have three independent constants,
but it can be less for various values of the parameters
$m_i^2$.  Substituting the solutions~\eqref{eq:2s-chi-soln} in the
remaining determining equation we find
\begin{equation}
  \alpha_1 m_1^2 \phi_1 + \alpha_2 m_2^2 \phi_2 + \beta (m_1^2-m_2^2) \phi_1\phi_2 = 0 \:.
\label{eq:2s-algdet}
\end{equation}
This final determining equation is now in an algebraic form.  It
is a polynomial in the fields, and each coefficient of the
polynomial must be independently zero.

Now comes the key observation: \emph{the symmetries depend on the
parameters}.  If $m_1^2=0$ then $\alpha_1$
is free.  Using the definition~\eqref{eq:eta-chi-sym}, we read off
the differential equations that this symmetry satisfies.  Setting
all other parameters to zero we obtain the differential equations
$\bar\phi_1'(\epsilon)=\alpha_1$ and $\bar\phi_2'(\epsilon)=0$.
Recall that $\epsilon$ is the parameter of the continuous group.
The solution is $\bar\phi_1(\epsilon)=\phi_1+\alpha_1\epsilon$ and
$\bar\phi_2(\epsilon)=\phi_2$.  We used the initial condition
$\bar\phi_i(0)=\phi_i$ to fix the integration constants.
Physically, this corresponds to a shift symmetry, and is only present
when $\phi_1$ is massless.
Indeed, if $m_1^2\ne0$ then equation~\eqref{eq:2s-algdet} is satisfied
only when $\alpha_1=0$.  Similarly, $\alpha_2$ is the shift
symmetry for $\phi_2$.

From equation~\eqref{eq:2s-algdet} we can also see that a symmetry
arises if $m_1^2=m_2^2$.  This symmetry has
$\bar\phi_1' = \beta \bar\phi_2$ and
$\bar\phi_2' = -\beta \bar\phi_1$.  The solution is
\begin{equation}
  \begin{pmatrix} \bar\phi_1 \\ \bar\phi_2 \end{pmatrix}
  =
  \begin{pmatrix} \cos \beta\epsilon & \sin \beta\epsilon \\
    -\sin \beta\epsilon & \cos \beta\epsilon \end{pmatrix}
  \begin{pmatrix} \phi_1 \\ \phi_2 \end{pmatrix} \:.
\end{equation}
This is a $\Guone$ rotation symmetry, and is only present when the
masses are equal.
$(\phi_1,\phi_2)\to(\bar\phi_1,\bar\phi_2)$ leaves the
action (and the Lagrangian in this case) invariant.

Each of the three independent symmetries we obtained has rank $R=(1)$
since they
correspond to a single parameter.  Depending on the values of the
masses, the total rank of the system will be different.  For
example, if $m_1^2=m_2^2\ne0$ then the total rank is $R=(1)$.  If
$m_1^2=m_2^2=0$ then the total rank is $R=(3)$.  It is important to
note that the rank, and the specific parameter relationships
leading to such a rank, can
be computed \emph{without} solving for the actual finite action of
the symmetries themselves.  Furthermore, the rank and parameter
relations can be obtain in a completely systematic way, as we
shall describe in Section~\ref{sec:auto}.  This is a crucial point
when analysing large systems.

That is the LPS method in a nutshell, from step 1, finding the
determining equations, through step 4, solving for the action of
the symmetries.  It seems simple, but when applying it to large
systems of equations one obtains orders of magnitude more determining
equations, along with much more interesting structure in the
symmetries.


\section{$N$ interacting scalar fields}
\label{sec:nsc}

The symmetries of a theory are dictated by the structure of the
interactions between fields, which are modelled by individual
terms in the action.  A generic theory with a certain set of
interactions will allow for a certain set of symmetry groups.  The
precise symmetries are then fixed by the specific values of the
coefficients of each interaction term.  This was made clear in the
previous section with the general form of the allowed set of
symmetries given by equation~\eqref{eq:2s-chi-soln}, while the
specific symmetries were dictated by the values of the two masses.
In general it is the terms with \emph{derivatives} in them that
are most important in dictating the allowed set of symmetries.
More complicated derivative terms allow for more sophisticated
symmetries.

In this section we use the LPS method to perform a complete
analysis of what is perhaps the simplest set of Lorentz invariant
interactions: the action corresponding to $N$ interacting spin-0
fields in $D$ dimensions.  The Lagrangian density is
\begin{equation}
  \mcl = T_{ij} \partial^\mu \phi_i \partial_\mu \phi_j - V(\phi) \:,
  \label{eq:nsc-lag}
\end{equation}
where $\phi_i$ are the $N$ fields with $i=1\ldots N$, $T_{ij}$ is
a constant $N\times N$ symmetric matrix describing the kinetic
mixing of the scalars, and $V(\phi)$ is an arbitrary potential of
all $N$ fields.  We work in flat space with mostly minus
signature, and a sum over repeated indices is implicit.
We can diagonalise $T_{ij}$, and then, assuming all its
eigenvalues are positive, rescale the fields to bring all the
kinetic terms to canonical form; in effect $T_{ij}\to\half\delta_{ij}$.
This will change the form of $V$, but since it is arbitrary this
makes no difference.

We now apply equation~\eqref{eq:actde} to this general Lagrangian,
in order to find the determining equations.  They are
\begin{align}
  V \partial_\mu \eta^\mu
    + \parpar{V}{\phi_i} \chi_i &= 0 \label{eq:nsc-eq1}\:,\\
  \partial^\mu \chi_i
    - V \parpar{\eta^\mu}{\phi_i} &= 0
    \qquad \forall \mu \; \forall i \label{eq:nsc-eq2}\:,\\
  \partial^\mu \eta^\nu + \partial^\nu \eta^\mu &= 0
    \qquad \forall \mu \; \forall \nu, \; \mu\ne\nu \:,\\
  \parpar{\chi_i}{\phi_j} + \parpar{\chi_j}{\phi_i} &= 0
    \qquad \forall i \; \forall j, \; i\ne j \:,\\
  \half \partial_\sigma \eta^\sigma
    - \partial_{\bar{\mu}} \eta^{\bar{\mu}}
    + \parpar{\chi_{\bar{i}}}{\phi_{\bar{i}}} &=0
    \qquad \forall \bar{\mu} \; \forall \bar{i} \label{eq:nsc-eq5}\:,\\
  \parpar{\eta^\mu}{\phi_i} &= 0
    \qquad \forall \mu \; \forall i \label{eq:nsc-eq6}\:.
\end{align}
Here a bar over an index indicates that an implicit sum should
not be taken.  Using equation~\eqref{eq:nsc-eq6} in
equation~\eqref{eq:nsc-eq2} we find that $\partial^\mu\chi_i=0$.
Thus, the coordinate transformations $\eta^\mu$ do not depend on
the fields, and the field transformations $\chi_i$ do not depend on
the coordinates.  This greatly simplifies the problem of solving
the above set of equations.

The general solution for $\chi_i$ is
\begin{equation}
  \chi_i(\phi) = \alpha_i + \beta_{ij} \phi_j + \gamma \phi_i \:,
  \label{eq:nsc-chi}
\end{equation}
where $\alpha_i$ and $\gamma$ are constants, and $\beta_{ij}$ is
antisymmetric: $\beta_{ij}=-\beta_{ji}$.  In terms of symmetries,
$\alpha_i$ corresponds to a constant shift of the fields,
$\beta_{ij}$ to rotations among the fields, and $\gamma$ to a
communal scaling symmetry.  These are, in general, the only symmetries
allowed in the field sector for the generic
Lagrangian~\eqref{eq:nsc-lag}, irrespective of the form of $V$.
We shall see later though that the constants in
equation~\eqref{eq:nsc-chi} are dictated by the choice of $V$
(or vice versa).

Let us now solve for $\eta^\mu$.  If we take
equation~\eqref{eq:nsc-eq5}, substitute in the solution for
$\chi_i$ and sum over $\bar{\mu}$ we obtain
\begin{equation}
  \left(\half D - 1 \right) \partial_\sigma \eta^\sigma + D \gamma = 0 \:.
  \label{eq:nsc-dbranch}
\end{equation}
At this point we get a branch in the solution, one for $D=2$ and
one for $D\ne2$.  Physically, this corresponds to the fact that
the scaling symmetry behaves differently in two dimensions because
the scalar field is dimensionless.

For the $D=2$ branch we have $\gamma=0$ and the general solution
for $\eta^\mu$ is
\begin{align}
  \eta^t &= F_+(t+x) + F_-(t-x) \:,\\
  \eta^x &= F_+(t+x) - F_-(t-x) + f \:,
\end{align}
where $F_+$ and $F_-$ are arbitrary functions of one variable, and
$f$ is a constant.  The final equation to solve is
equation~\eqref{eq:nsc-eq1}:
\begin{equation}
  2\left[F'_+(t+x) + F'_-(t-x)\right] V
    + \parpar{V}{\phi_i} \left( \alpha_i + \beta_{ij} \phi_j \right) = 0 \:.
  \label{eq:nsc-d2det}
\end{equation}
Since there are the free constants $f$, $\alpha_i$ and
$\beta_{ij}$, and the two free functions $F_{\pm}$, the maximum
rank of this system is $R=(1+N+N(N-1)/2,2)$.

For $D\ne2$ we can solve equation~\eqref{eq:nsc-dbranch} for
$\partial_\sigma \eta^\sigma$ and then proceed to determine the
general solution for $\eta^\mu$.  It is
\begin{equation}
  \eta^\mu(x) = a^\mu
    + b^\mu_{\phantom{\mu}\nu} x^\nu
    - \frac{2\gamma}{D-2} x^\mu \:.
\label{eq:nsc-eta}
\end{equation}
Here, $a^\mu$ corresponding to coordinate translations,
$b^{\mu\nu}$ is antisymmetric and corresponds to
coordinate rotations and boosts, and $\gamma$ gives the
scaling of the coordinates.  Note the similarities between this
equation and equation~\eqref{eq:nsc-chi}.  Also note that $\gamma$
is the only symmetry parameter connecting the field and coordinate
sectors.  The maximum rank of the system is the number of free
constants in the general solutions~\eqref{eq:nsc-chi}
and~\eqref{eq:nsc-eta}, which is $R=(D+D(D-1)/2+1+N+N(N-1)/2)$.
The remaining equation to solve for this $D\ne2$ branch is
\begin{equation}
  -d\gamma V
    + \parpar{V}{\phi_i}
      \left( \alpha_i + \beta_{ij} \phi_j + \gamma \phi_i \right) = 0 \:,
  \label{eq:nsc-dn2det}
\end{equation}
where we defined $d\equiv2D/(D-2)$.

We have now reduced the original set of determining equations to
equation~\eqref{eq:nsc-d2det} for $D=2$, and
equation~\eqref{eq:nsc-dn2det} for $D\ne2$.  We have also solved
for the generic form of $\eta^\mu$ and $\chi_i$.  Further progress
can be made if one chooses a specific
$N$ and/or $V$.  Once these are specified, one can continue to
find any relationships between the symmetry parameters $\alpha_i$,
$\beta_{ij}$, $\gamma$, $F_+$ and $F_-$.  Alternatively, one can
specify these parameters~--- hence specify a symmetry~--- and look
for a potential $V$ that allows for this.  In what follows we
focus on a few of the simpler cases.


\subsection{The case $D=2$ and $N=1$}
\label{sec:nsc-d2n1}

In two dimensions with one field we have the general solution
$\chi=\alpha$, which is a shift symmetry.  Depending on the form
of $V$, $\alpha$ may be restricted.  The final determining
equation~\eqref{eq:nsc-d2det} becomes
\begin{equation}
  2 X(t,x) V + \frac{\md V}{\md \phi} \alpha = 0 \:,
\label{eq:nsc-d2n1det}
\end{equation}
where $X(t,x)=F'_+(t+x) + F'_-(t-x)$.  If $X(t,x)$ is not a
constant then the only solution is $V=0$, and the symmetry rank is
$R=(2,2)$.  If $X$ is a constant (or zero) then we can solve for
$F_{\pm}$ and then obtain the $\eta$'s:
\begin{align}
  \eta^t &= a^t + (b-X)x + Xt \:,\\
  \eta^x &= a^x + (b-X)t + Xx \:.
\end{align}
Since $b$ is free we can redefine $b\to b+X$ to bring these
solutions to canonical form.  Then the symmetries corresponding to
the constants $a^t$, $a^x$, $b$ and $X$ are, respectively, time
translations, space translations, boosts and scaling.

For $X$ a constant there are a few different cases to consider
depending on the form of $V(\phi)$.  If $V=0$ then $X$ does not
need to be a constant, and we have already considered this case.
If $V$ is a constant then $X=0$ but $\alpha$ is free, and we have
a total rank $R=(4)$.  For a non-trivial solution we solve the
differential equation~\eqref{eq:nsc-d2n1det} for $V$ to obtain
\begin{equation}
  V = \lambda \me^{-\frac{2X}{\alpha}\phi} \:.
\label{eq:nsc-d2n1v}
\end{equation}
If $V$ is of this form then $X$ and $\alpha$ are non-zero and are
related given a certain choice for $V$.  In this case the scale
symmetry must work in combination with a shift of the field.  The
total rank for this case is $R=(4)$.  Finally, for an arbitrary
potential $V$ that is non-zero, is not a constant and does not
have the form of equation~\eqref{eq:nsc-d2n1v}, the constants $X$
and $\alpha$ must be zero and the total rank of the symmetry for
this generic case is $R=(3)$.

In summary, for $D=2$ and $N=1$ there are the following distinct
sets of symmetries:
\begin{caselist}
\item[$V=0$:] $X(t,x)=F_+'(t+x)-F_-'(t-x)$ can take any form and
$f$ and $\alpha$ are also free.  The rank is $R=(2,2)$.
\item[$V=\text{const}$:] $X=0$, but $a^{t,x}$, $b$ and $\alpha$
are free.  The rank is $R=(4)$.
\item[$V=\lambda \me^{-m\phi}$:] $X$ and $\alpha$ are related by
$m=2X/\alpha$, so the scale symmetry is combined with a shift.
The parameters $a^{t,x}$ and $b$ are free.  The rank is $R=(4)$.
\item[$V$ arbitrary:] $X=\alpha=0$, but $a^{t,x}$ and $b$ are free.
The rank is $R=(3)$.
\end{caselist}


\subsection{The case $D\ne2$ and $N=1$}
\label{sec:nsc-dn2n1}

We now consider one scalar field in dimensions other than two.
Equation~\eqref{eq:nsc-dn2det} becomes
\begin{equation}
  -d\gamma V
    + \frac{\md V}{\md \phi} \left( \alpha + \gamma \phi \right) = 0 \:.
  \label{eq:nsc-dn2n1det}
\end{equation}
There are two parameters in the field sector, $\alpha$ and
$\gamma$.  As in the $D=2$ case we get four distinct cases which
depend on the form of the potential:
\begin{caselist}
\item[$V=0$:] $\alpha$ and $\gamma$ free, so there exist
independent shift and scale symmetries.  Rank associated with the
field is $R_\chi=(2)$.
\item[$V=\text{const}$:] $\gamma=0$ but $\alpha$ is free.
Field rank $R_\chi=(1)$.
\item[$V=\lambda(\phi+v)^d$:] This form of $V$ is obtained by
solving the differential equation~\eqref{eq:nsc-dn2n1det}.  Given
a specific value of $v$, the relationship between the shift and
scale symmetry is then fixed by $v=\alpha/\gamma$.  Note that if
$v=0$ then the theory has the usual scale symmetry, otherwise it
is a combined shift-scale symmetry.
The field rank is $R_\chi=(1)$.
\item[$V$ arbitrary:] $\alpha=\gamma=0$, so no shift or scale
symmetry. Field rank $R_\chi=(0)$.
\end{caselist}

Recall that $\eta^\mu$ has solution~\eqref{eq:nsc-eta}, which
includes coordinate shifts and rotations/boosts, along with a
scaling symmetry if $\gamma\ne0$.  The total rank includes the
rank from the coordinates:
$R_{\text{total}}=R_\chi+R_\eta=R_\chi+(D+D(D-1)/2)$.


\subsection{The case $D\ne2$ and $N=2$}

Moving on to two scalar fields, in dimensions other than two, the
remaining determining equation is
\begin{equation}
  -d\gamma V
    + \parpar{V}{\phi_1}\left(\alpha_1 + \beta\phi_2 + \gamma\phi_1\right)
    + \parpar{V}{\phi_2}\left(\alpha_2 - \beta\phi_1 + \gamma\phi_2\right)
    = 0 \:.
\label{eq:nsc-dn2n2det}
\end{equation}
Here we defined $\beta\equiv\beta_{12}$.  We cannot solve this in
a general way like we did in the previous cases with $N=1$.  One
option is to specify a particular form for $V$ (like a polynomial)
then solve for the symmetries.  Alternatively, try to find a $V$
that yields a given symmetry.  As seen in the $N=1$ cases,
it is possible to obtain non-conventional single
parameter symmetries which are compositions of more familiar
symmetries such as shifting and scaling.  For the $N=2$ case there
is the potential to have a relation between $\beta$ and $\alpha$
or $\beta$ and $\gamma$, yielding, respectively, combined
shift-rotation and combined scale-rotation symmetries.

Actually, if we go to polar field variables then we can make
some progress with equation~\eqref{eq:nsc-dn2n2det}.  The general
polar Lagrangian density is
\begin{equation}
  \mcl =
    \half \partial^\mu r \partial_\mu r
    + r^2 \half \partial^\mu \theta \partial_\mu \theta
    - V(r,\theta) \:.
\label{eq:nsc-polar-lag}
\end{equation}
This has the usual Poincar\'e symmetry, and possibly scaling, with
the general solution for $\eta^\mu$ given by
equation~\eqref{eq:nsc-eta}.  The polar version of the determining
equation~\eqref{eq:nsc-dn2n2det} is\footnote{One can either
transform equation~\eqref{eq:nsc-dn2n2det} directly to polar form,
or rederive it from scratch in the polar basis using the
Lagrangian~\eqref{eq:nsc-polar-lag}.  The result is the same.}
\begin{equation}
  -d\gamma V
    + \parpar{V}{r}\left(\alpha_1\cos\theta + \alpha_2\sin\theta + \gamma r\right)
    + \parpar{V}{\theta}\left(-\alpha_1\frac{\sin\theta}{r} + \alpha_2\frac{\cos\theta}{r} - \beta \right) = 0 \:.
\end{equation}
Note that if $V$ does not depend on $\theta$ then $\beta$ is free
and there is a rotation symmetry (which manifests as a shift of
the $\theta$ field).  Furthermore, $\alpha_{1,2}=0$ and the
general solution for the potential is $V(r)=\lambda r^d$.  Such a
potential has a rotation and scale symmetry, but they act
independently.

If $V$ does depend on $\theta$ but now not on $r$ then we find that
$\alpha_{1,2}=0$ and we are left with the differential equation
\begin{equation}
  d \gamma V + \frac{\md V}{\md \theta}\beta = 0 \:.
\end{equation}
The solution for the potential is
$V(\theta)=\lambda \exp(-d\gamma\theta/\beta)$.
If the potential has this form then $\gamma/\beta$ is fixed and
the symmetry acts by scaling $x^\mu$ and $r$ in combination with a
shift of $\theta$.  This is in effect a combined scale-rotation
symmetry, or a spiral symmetry.\footnote{The spiral symmetry acts
on the polar Lagrangian~\eqref{eq:nsc-polar-lag}.  Due to the
multivalued nature of the inverse tangent function, it is
difficult to define the corresponding spiral symmetry in Cartesian
field variables.}  We can also obtain other spirals.  For example,
with $\alpha_{1,2}=0$, the following will work:
\begin{equation}
  V(r,\theta) = \lambda\left(r^k-v \me^{l\theta}\right)^m \:.
\label{eq:nsc-spiral-v}
\end{equation}
The parameters $\lambda$, $v$ and $l$ are free, while $k$ and $m$
are related by $mk=d$ (for example, for $D=4$ one can choose
$k=m=2$).  The relationship between the scale and
rotation symmetry is fixed by $k\gamma=l\beta$.  The action of the
symmetry is $r\to \me^\gamma r$, $\theta\to\theta-k\gamma/l$ and
$x^\mu\to \me^{-d\gamma/D}x^\mu$, which is parameterised here by
$\gamma$.  For small $l$ and/or $\theta$ one can Taylor expand the
exponential in equation~\eqref{eq:nsc-spiral-v} to obtain a
polynomial potential which is, up to higher order corrections in
$l$ and/or $\theta$, spirally symmetric.  It would be interesting
to see if this result extends to a non-Abelian rotation.


\section{The action versus the equations of motion}
\label{sec:act-vs-eom}

There is a distinction between the symmetries of an action and the
symmetries of the corresponding set of equations of motion, the
Euler-Lagrange equations.  Indeed, if $G$ is a symmetry of a given
action then $G$ is also a symmetry of the Euler-Lagrange
equations, but the converse is not necessarily true.\footnote{See
Section 4.2 of Olver's book~\cite{Olver:1986aa}.}  The additional
symmetries present in the Euler-Lagrange equations are physically
meaningful.  We shall illustrate these points with an example,
which also serves to illustrate the alternative way of
obtaining the determining equations as described in
Section~\ref{sec:lps-eom}.  While this alternative method in
general requires a lot more effort, it is applicable to any set
of differential equations, not just those arising from a field
theory.  The derivation in this section also shows in more detail
how one can determine the symmetry rank of a system without
actually solving for the action of the symmetry, something which
is important for automated solving of large systems.

Consider a single massive real scalar in 2d, whose Euler-Lagrange
equation is $\ddot{\phi} - \phi'' + m^2 \phi=0$.  As per
Section~\ref{sec:lps-eom}, the symmetry operator is
$\symop=\eta^t\partial_t+\eta^x\partial_x+\chi\partial_\phi$.
We apply the second prolongation of this operator,
$\pr^{(2)}\symop$, to the equation of motion, use the equation of
motion to eliminate $\ddot{\phi}$ (alternatively $\phi''$), and
then equate the coefficients of all independent derivatives of the
field to zero.  This results in 27 raw determining equations.
Only 14 of these are unique (read the following as all
individually equated to zero):
\begin{equation}
\begin{aligned}
& \eta^t_\phi \:;\qquad
  \eta^t_{\phi\phi} \:;\qquad
  \eta^t_{t\phi} \:;\qquad
  \eta^x_\phi \:;\qquad
  \eta^x_{\phi\phi} \:;\qquad
  \eta^x_{t\phi} \:;\\
& \eta^t_t - \eta^x_x \:;\qquad
  \eta^t_x - \eta^x_t \:;\qquad
  \eta^t_{x\phi} - \eta^x_{t\phi} \:;\qquad
  2 \eta^t_{t\phi} - \chi_{\phi\phi} \:;\qquad
  2 \eta^x_{x\phi} - \chi_{\phi\phi} \:;\\
& \eta^t_{tt} - \eta^t_{xx} - 3m^2 \phi \eta^t_\phi - 2 \chi_{t\phi} \:;\qquad
  \eta^x_{tt} - \eta^x_{xx} - m^2 \phi \eta^x_\phi + 2 \chi_{x\phi} \:;\\
& \chi_{tt} - \chi_{xx} + m^2 \chi - m^2 \phi \chi_\phi + 2m^2 \eta^t_t \:.
\end{aligned}
\end{equation}
The notation is that a subscript denotes differentiation with
respect to that variable.
The aim is to solve this set of equations for the functions
$\eta^t(t,x,\phi)$, $\eta^x(t,x,\phi)$ and $\chi(t,x,\phi)$.
Simple and obvious substitutions reduce this set to eight
equations.  Then integrability conditions (basically taking
derivatives and linear combinations of the above) can simplify the
system further, although we end up with more equations (10 of
them, but one is actually redundant):
\begin{equation}
\begin{aligned}
& \eta^t_\phi \:;\qquad
  \eta^x_\phi \:;\qquad
  \chi_{t\phi} \:;\qquad
  \chi_{x\phi} \:;\qquad
  \chi_{\phi\phi} \:;\\
& \eta^t_x - \eta^x_t \:;\qquad
  \eta^t_t - \eta^x_x \:;\qquad
  \eta^t_{tt} - \eta^t_{xx} \:;\qquad
  \eta^x_{tt} - \eta^x_{xx} \:;\\
& \chi_{tt} - \chi_{xx} + m^2 \chi - m^2 \phi \chi_\phi + 2m^2 \phi \eta^t_t \:.
\end{aligned}
\label{eq:1fs-b4branch}
\end{equation}
Taking the derivative of the last equation with respect to $\phi$
and using the other equations to make eliminations one finds that
$m^2 \eta^t_t = 0$.  At this point we must create two branches of
possible solutions, one when $m=0$ and one when $m\ne0$.  Such a
branch is a key step of the LPS method.  In more complicated cases
it leads to more interesting relationships between parameters.
When automating the procedure, this is the point at which the
algorithm must also branch.

\subsection{The $m=0$ branch}

Take $m=0$ to begin with.  Instead of writing down trivial first
derivative equations like $\eta^t_\phi=0$ we shall just solve this
equation and redefine the function to not depend on that
particular variable.  This branch then has the functions
\begin{equation}
  \eta^t(t,x) \:,\qquad
  \eta^x(t,x) \:,\qquad
  \chi(t,x,\phi) \:,
\end{equation}
with determining equations
\begin{equation}
\begin{aligned}
& \chi_{t\phi} \:;\qquad
  \chi_{x\phi} \:;\qquad
  \chi_{\phi\phi} \:;\\
& \eta^t_x - \eta^x_t \:;\qquad
  \eta^t_t - \eta^x_x \:;\qquad
  \eta^t_{tt} - \eta^t_{xx} \:;\qquad
  \eta^x_{tt} - \eta^x_{xx} \:;\qquad
  \chi_{tt} - \chi_{xx} \:.
\end{aligned}
\label{eq:1fs-meq0-deteqns}
\end{equation}
This is as far as one needs to go to determine the rank of the
symmetry group for this branch (the above set of PDEs is in
involution).  The rank is determined by working out how much
initial data one needs in order to fully specify a unique
solution to the above system.  This initial data is computed, via
a well defined algorithm~\cite{Reid:1990aa}, to be the values of
\begin{equation}
  \eta^t(0,0) \:,\qquad
  \eta^x(0,x) \:,\qquad
  \eta^x_t(0,x) \:,\qquad
  \chi(0,x,0) \:,\qquad
  \chi_t(0,x,0) \:,\qquad
  \chi_\phi(0,0,0) \:.
\end{equation}
There are two constants, $\eta^t(0,0)$ and $\chi_\phi(0,0,0)$, and
four functions of one variable.  We do not need to evaluate the
coordinates and field at $0$, any arbitrary value would do to fix
the solution.  All that matters is how much data is needed.  In
this case the rank of the symmetry group is $R=(2,4)$, with the
$\eta$'s contributing $R_\eta=(1,2)$ and $\chi$, $R_\chi=(1,2)$.

If all that is needed is the rank of the symmetry for a particular
branch then one can stop here.  But we shall solve the determining
equations fully to show exactly what the symmetries of the system
are and how they differ from the analysis performed with the
action approach.

The general solution of the set of
equations~\eqref{eq:1fs-meq0-deteqns} is
\begin{equation}
\begin{aligned}
  \eta^t(t,x) &= F_+(t+x) + F_-(t-x) \:,\\
  \eta^x(t,x) &= F_+(t+x) - F_-(t-x) + f \:,\\
  \chi(t,x,\phi) &= G_+(t+x) + G_-(t-x) + g\, \phi(t,x) \:.
\end{aligned}
\end{equation}
As previously calculated via the initial data, we have two free
constants, $f$ and $g$, and four free functions of one variable,
$F_+$, $F_-$, $G_+$ and $G_-$.  Each one of these corresponds to
an independent continuous symmetry of the original Euler-Lagrange
equation (with $m=0$).  $f$ is a spatial translation and $g$ is a
scaling of $\phi$ (without scaling the coordinates).

For the functions $G_\pm$ the equations describing the action
of the symmetry are $\bar{t}'=0$, $\bar{x}'=0$ and
$\bar{\phi}' = G_\pm(\bar{x}\pm\bar{t})$.  The solution is
$\bar{t}=t$, $\bar{x}=x$ and
$\bar{\phi} = \phi+\epsilon G_\pm(t\pm x)$.
Therefore, given a solution $\phi(t,x)$ of the equation of motion,
one can add an arbitrary function $G_\pm(t\pm x)$ to that solution
and the result is still a solution.  Thus $G_\pm$ corresponds to
additivity/superposition of solutions of $\phi(t,x)$.  (Note that
the massless field $\phi$ has the general wave solution
$\phi=w(t\pm x)$ with $w$ arbitrary.)  The symmetries due to the
functions $F_\pm$ are difficult to solve for in general, but
include the Poincar\'e group and coordinate scaling.  For example,
choosing $F_+=F_-=\text{constant}$ yields temporal translations,
while choosing $F_\pm(t\pm x)=t \pm x$ yields a scaling symmetry.

Note that $x$-translations correspond to a distinct piece of
initial data $f$, whereas $t$-translations do not.  There is
nothing important behind this asymmetry; it is simply because of
the way we solved~\eqref{eq:1fs-meq0-deteqns}, choosing to solve
first for $\eta^t$ and then for $\eta^x$.  It is important to
remark that the derived parameter relationships and the
computation of the rank is independent of the way in which the
determining equations are reduced and solved.

\subsection{The $m\ne0$ branch}

For the other branch with $m\ne0$ we take the set of
equations~\eqref{eq:1fs-b4branch} and solve it for $\eta^t_t=0$.
We get the functional dependence
\begin{equation}
  \eta^t(x) \:,\qquad
  \eta^x(t) \:,\qquad
  \chi(t,x,\phi) \:,
\end{equation}
and determining equations
\begin{equation}
\begin{aligned}
& \chi_{t\phi} \:;\qquad
  \chi_{x\phi} \:;\qquad
  \chi_{\phi\phi} \:;\qquad
  \eta^t_{xx} \:;\qquad
  \eta^x_{tt} \:;\qquad
  \eta^t_x - \eta^x_t \:;\\
& \chi_{tt} - \chi_{xx} + m^2 \chi - m^2 \phi \chi_\phi \:.
\end{aligned}
\label{eq:1fs-mne0-deteqns}
\end{equation}
From here the rank can be determined.  The initial data consists
of the values of
\begin{equation}
\eta^t(0) \:,\qquad
\eta^x(0) \:,\qquad
\eta^x_t(0) \:,\qquad
\chi(0,x,0) \:,\qquad
\chi_t(0,x,0) \:,\qquad
\chi_\phi(0,0,0) \:.
\end{equation}
The rank is $R_\eta=(3,0)$ and $R_\chi=(1,2)$,
giving a total rank of $R=(4,2)$.  We can see that the symmetry is
less than the $m=0$ case (as expected), and that the symmetry
reduction is in the coordinate sector rather than the field
sector.

The general solution of the determining
equations~\eqref{eq:1fs-mne0-deteqns} is
\begin{equation}
\begin{aligned}
  \eta^t(x) &= a^t + b x \:,\\
  \eta^x(t) &= a^x + b t \:,\\
  \chi(t,x,\phi) &= \int_{-\infty}^{+\infty} dk
    \left[ H_+(k)\, \me^{i(\omega t + k x)}
      + H_-(k)\, \me^{i(\omega t - k x)} \right]
    + g\, \phi(t,x) \:,
\end{aligned}
\end{equation}
where $\omega=\sqrt{k^2+m^2}$.  In the coordinate sector there is
exactly the Poincar\'e group.  In the field sector there is
scaling, and the $H_\pm(k)$ correspond to additivity of backward
and forward waves with wave number $k$.

This completes our analysis of a single scalar in 2d using the
Euler-Lagrange approach.  Compare with the analysis using the
action approach in Section~\ref{sec:nsc-d2n1}: there we found rank
$R=(2,2)$ for the massless case and $R=(3,0)$ for the massive
case, compared to here with $R=(2,4)$ and $R=(4,2)$ respectively.
The main difference is that the equations of motion are invariant
under addition of solutions, whereas the action is not.  An
important question, which we shall not attempt to answer, is
whether the larger class of symmetries obtained from the
Euler-Lagrange equations contains anything interesting and/or
important when analysing large systems.  From a pragmatic point of
view the action approach is much simpler (it yields fewer
determining equations) and this approach would be preferred for
large systems.\footnote{We could also consider a third class of
symmetries, those of the Lagrangian density itself (not the
action).  This is probably not very interesting once we have the
symmetries of the action and/or Euler-Lagrangian equations, but it
is easy to do.  We just take equation~\eqref{eq:actde} and drop
the first term.  The resulting equation can give the determining
equations for the pure Lagrangian (without the volume element).}


\section{Remarks on the LPS method}
\label{sec:remarks}

Of great interest is the fact that the LPS method provides an
exhaustive list of relationships between parameters such that an
enhanced symmetry is obtained.
In the reduction and solving of the determining equations one
invariably comes across equations where different solutions are
obtained (and hence different symmetries) when parameters of the
model take special values, or when unspecified functions like the
potential take a different functional form.\footnote{Unspecified
functions can only be functions of $x$ and $\phi$, not of
derivatives of $\phi$, since then one cannot obtain the
determining equations in explicit form.}
The converse is also
true, because if special values or combinations of parameters, or
special forms of unspecified functions, lead to different
symmetries then the LPS method must necessarily distinguish these
scenarios.  In the previous sections we have demonstrated this for
some simple systems.  For more complex cases one can obtain much
more complicated relationships among the parameters.

In the examples so far we have only considered spin-0 particles,
and the LPS method has been developed to handle only a collection of
real fields.  But in fact any spin representation, or even particles
that do not respect Lorentz symmetry, can be written in terms of
real fields, and, consequently, any action can be
expanded in terms of its real components.  The LPS method, and in
particular equation~\eqref{eq:actde}, is therefore general enough
for the purposes of model building.  For example, in 4d, gauge
fields and Weyl fermions have four real components.  From this
point of view an action is just a bunch of real fields with
certain terms being derivative interactions and other terms having
the usual Yukawa form.  The distinguishing feature between
scalars, gauge fields and fermions is the structure of the
self-coupling derivative (kinetic) terms.  For a fermion such
derivative terms are just right to get a spin-$\half$ representation
of the Lorentz group.  Any field theory is then just a set of
interacting real fields, and the symmetries of the theory act on
this set, and on the coordinates.  This is a very na\"ive view,
but with this na\"ivety comes freedom from bias and allows one to
systematically classify and study the properties of a given model.
We return to this point in Section~\ref{sec:auto}.

The LPS method works for all continuous symmetries that depend on
the coordinates and fields (but not derivatives of the fields).
This includes local gauge symmetries~\cite{Hereman:1993aa,Marchildon:1995ma}
as well as local coordinate diffeomorphisms~\cite{Marchildon:1995ma}.
The extension to supersymmetry is possible and requires the
introduction of anti-commuting coordinates~\cite{Grundland:2008ak}.
The LPS method works also for non-linear symmetries.  As an
example, consider the field (no coordinate) symmetries of
$\mcl = \phi^m (\partial^\mu \phi \partial_\mu \phi)^n$, where $m$
and $n\ne0$ are constant exponents.  The solution to the
determining equations is $\chi = a \phi^{-m/2n}$ with constant
$a$.  The corresponding non-linear symmetry acts by
$\phi\to(\phi^p+pa\epsilon)^{1/p}$, with $p=1+m/2n$.

Section~\ref{sec:nsc-dn2n1} gave a simple example where a
spontaneously broken scale symmetry was found, with potential
$V=\lambda(\phi+v)^d$.  The corresponding unbroken potential is
$V=\lambda\phi^d$, which is manifestly scale invariant.  The
potential written in the broken phase still has the same symmetry
rank, but now the particular scale symmetry is implemented by a
combined shift-scale.  The symmetry acts by first shifting the
field to the unbroken phase, then scaling it, then shifting it
back to the broken phase.  This is a generic feature of
spontaneously broken symmetries.  A Lagrangian that is symmetric
under a certain group, which is then expanded around some vacuum
state, will still contain essentially the same symmetry, just
implemented in a slightly different way.  At the level of the
action, one cannot break a symmetry by a simple field
redefinition (it is the vacuum state, hence a solution of the
equations of motion, that breaks the symmetry).  This means that
the LPS method will always be able to find a symmetry of an
action, even if the action is written in the broken phase.

In Section~\ref{sec:nsc} we analysed $N$ interacting spin-0
fields, and proved that the most general symmetries of such
a system are given by equations~\eqref{eq:nsc-chi}
and~\eqref{eq:nsc-eta} (for dimensions other than 2d).  The LPS
method thus gives a complementary approach to the proofs regarding
all symmetries of the
S-matrix~\cite{Coleman:1967ad, Haag:1974qh}.  The method at
hand also gives a straightforward way to compute the precise
symmetries for a given model.  Furthermore, as pointed out above,
it is applicable to spontaneously broken symmetries.

Looking beyond point symmetries, one has contact symmetries, which depend also
on the first derivative of the field, and generalised symmetries
(sometimes called Lie-B\"acklund), which allow $\eta^\mu$ and
$\chi_i$ to depend on arbitrary derivatives of $\phi_i$.  These
can be handled with suitable extensions of the LPS method,
although solving the determining equations becomes more
involved.  There are also discrete symmetries, which, apart from
those that are subsets of a continuous group, are not covered by
the methods outlined in this paper.  One can easily show that
continuous and discrete symmetries act independently: if
$g\to C(D(g))$ is a symmetry then so is $g\to C(g)$ and
$g\to D(g)$, for $C$ a continuous, and $D$ a discrete symmetry.
Thus, if we find all the continuous symmetries, and then all
discrete symmetries separately, then we have found all of the
symmetries of the system.  Ref.~\cite{Hydon:2000aa} discusses a
method to systematically find discrete point symmetries, which
should be applicable to model building.  See also
Ref.~\cite{Low:2003dz} for a systematic study of discrete
symmetries in the context of model building.


\section{Automation, and a catalogue of all field theories}
\label{sec:auto}

Applying the LPS method to a large, complex system can lead to an
unmanageable set of determining equations.  Fortunately, the
procedure can be cast as a well defined algorithm that completes
in finite time, at least up to finding parameter relationships and
the rank of the symmetry.  It is therefore feasible to construct a
computer program which takes in a Lagrangian and returns a list of
branches, where each branch corresponds to a different set of
symmetries and consists of the associated rank and parameter
relationships.  In addition the branch can contain the reduced
determining equations, which can be further solved if needed.
We shall outline how such a program can be constructed, and then
present a few ideas on how it can be put to use.

First of all the program must compute the determining equations.
This is straightforward using standard computer algebra, although
for systems with a large number of degrees of freedom one must be
careful to use symbolic algebra algorithms that have low order
complexity in the number of terms.

Reducing the determining equations to standard form is the difficult
part.  Let us first discuss how this works for the simplified case
where the determining equations are reduced to algebraic form, as
happened in the example with two massive scalars in the derivation
of equation~\eqref{eq:2s-algdet}.  From that equation one obtains
three independent constraints which can be written in matrix form
as
\begin{equation}
  \begin{pmatrix} m_1^2 & 0 & 0 \\ 0 & m_2^2 & 0 \\ 0 & 0 & m_1^2-m_2^2 \end{pmatrix}
  \begin{pmatrix} \alpha_1 \\ \alpha_2 \\ \beta \end{pmatrix} = 0 \:.
\end{equation}
Any set of algebraic determining equations can be written this
way, with the matrix containing the free parameters of the theory, and
the column vector the symmetry parameters.  We are interested in this
matrix's null space, which can be different for special values of
the model parameters.  In the above example, if the masses are not
special then the null space is trivial,
$(\alpha_1,\alpha_2,\beta)=0$ and there are no symmetries.  But, if
$m_1^2=0$, $m_2^2=0$ or $m_1^2=m_2^2$, then one of the rows of the
matrix is eliminated and the dimension of the null space is at
least one, meaning there is at least on independently acting
symmetry.  For large systems the matrix can be large and contain
off diagonal entries, and there can be many cases to consider.
The problem can be solved
systematically by implementing a Gaussian elimination algorithm
that takes into account the possibility of a leading entry being
zero or non-zero, and producing a new branch at such a point.

For the general case the determining equations are linear
partial differential equations in $\eta^\mu$ and $\chi_i$, and
one reduces the system to ``diagonal'' form using a
generalised version of Gaussian elimination.  For this to make
sense, one first defines a strict ordering of the variables
$\eta^\mu$ and $\chi_i$ and their derivatives, with higher order
derivatives coming first.  The terms in each determining equation
are then sorted using this ordering.  In principle this linear
system can then be written as a matrix of coefficients (which can
depend on $x^\mu$ and $\phi_i$) operating on a vector of all
possible derivatives of $\eta^\mu$ and $\chi_i$.  Generalised
Gaussian elimination can then proceed, with additional operations
such as differentiation of a matrix row.  In practice the matrix
is very sparse and it is easier to implement the rules of
reduction directly on the determining equations.

It is during the equivalent of column elimination that branching
of the solution can occur.  Schematically, the leading order terms
in a pair of determining equations looks like
\begin{equation}
\begin{aligned}
  c_1\, \partial_i f + X_1(f) &= 0 \:,\\
  c_2\, \partial_{i+j} f + X_2(f) &= 0 \:.
\end{aligned}
\end{equation}
Since the derivative $\partial_i$ is contained within
$\partial_{i+j}$ we can use the first equation to eliminate the
$c_2\, \partial_{i+j} f$ term in the second equation.  But this can
be done only if the leading coefficient $c_1$ is non-zero.  Since
$c_1$ in general depends on free parameters, or free functions, of
the original model, $c_1$ being zero or non-zero defines a
particular branch point in the solution, and a particular
relationship between parameters that may lead to a different set
of symmetries.  Each branch is reduced until no more eliminations
can be done, at which point the system is in involution and
includes all of its integrability conditions.  In this
way the determining equations are systematically reduced, with all
branching accounted for.  See Reid~\cite{Reid:1990aa,Reid:1991aa}
for a more detailed description of this algorithm, and also for an
algorithm which computes the initial data, the rank, of the
reduced set of determining equations.

Let us now assume that we have a program which, given a
Lagrangian, can tell us in a reasonable amount of time all the
possible branches and their corresponding parameter constraints
and symmetry rank (and possibly also the symmetry group).  Now,
using the observation that any model can be written in terms of
its real components, we can start to make a comprehensive
catalogue of all possible theories, at least within some limit.

Such a catalogue will be ordered on the number of real degrees of
freedom $N$.  Given this number, we literally just write down the
most general action, with general derivative couplings and Yukawa
couplings between all $N$ degrees of freedom.  Feeding the action
into our program we obtain a large but finite list of all the
possible relationships among parameters and all the symmetries.
This will include relationships amongst the coefficients of the
derivative terms in order to get particles of spin-0, spin-$\half$
and so on.  For a given $N$, given number of dimensions and given
highest-order coupling, this is a well defined and finite
procedure.  Any model one can think of will be in this catalogue.

For instance, in 4d with $N=4$ one will find electromagnetism,
$N=6$ contains scalar QED, and $N=8$ has QED with one Weyl
fermion.  With $N=10$ one will find general relativity, among many
other theories.  It may seem that one will only ``find'' general
relativity because one had prior knowledge of what to look for,
but it is arguable that the symmetry rank of general relativity is
so much larger than others in the $N=10$ class that it would stand
out from the rest of the branches.  If true, simply by sorting the
branches on their rank would allow one to literally ``discover''
general coordinate invariance.

As an explicit example, with $N=1$ in 2d with up to bi-linear
terms we would write
\begin{equation}
\mcs = \int \md t \, \md x \left[
  c_0
  + c_1 \phi
  + c_2 \dot{\phi}
  + c_3 \phi'
  + c_4 \phi^2
  + c_5 \phi \dot{\phi}
  + c_6 \phi \phi'
  + c_7 \dot{\phi}^2
  + c_8 \dot{\phi} \phi'
  + c_9 \phi'^2
\right] \:,
\end{equation}
where the $c_i$ are constant parameters of the theory.  Feeding
this into our program would give us all the relationships among
the $c_i$ along with the corresponding symmetries.  It would
include the case $c_0=c_1=c_2=c_3=c_5=c_6=c_8=0$, $c_7=-c_9$ which
is a massive spin-0 field in 2d.

The obvious drawback of all this is the computational limits in
time and storage that will be hit for moderately sized systems.
The reason is that the number of terms in a general action grows
combinatorically with the number of fields, and even more so with
the number of coordinates.  For $D$ coordinates, $N$ fields and
a maximum of $F$ factors in each term, there are
\begin{equation}
T = \frac{(F + N + DN)!}{F!(N + DN)!}
  = \frac{(F + N + DN) \cdots (1 + N + DN)}{F!}
\end{equation}
distinct terms.  For $D=F=2$ we have $T=(9N^2+9N+2)/2$ which
behaves asymptotically like $N^2$.  For $D=F=4$ (4d with
renormalisable terms) we obtain $T\sim26N^4$ for large $N$,
and for $N=10$ there are about $3\times10^5$ terms (and this many
free parameters), which may be manageable by a computer.  At least
for $N=4$ it will most likely be manageable with $T\sim10^4$,
allowing for a gauge field or a Weyl fermion.

To ease the combinatorical problem we can restrict the derivative
terms to a known spin structure.  The class of models to be
considered is then designated by the number of spin-0 fields,
number of spin-$\half$ fields (Weyl) and so on, and the number of
parameters in the initial action.  For each model we obtain a list
of all the branches of possible sets of symmetries, with the
number of remaining free parameters, and the rank of the symmetry
group.  The classification label for a model $\mcm$ might look
something like
\begin{equation}
\mcm =
  (N_\text{spin-0},N_\text{spin-$\half$},N_\text{spin-1},N_\text{param})
  \to \left\{
    (N_\text{free param},R_\eta,R_\chi)
  \right\} \:,
\label{eq:mclass}
\end{equation}
where $N_X$ is the number of $X$ and $R$ is the symmetry rank.
For a single spin-0 particle in 2d with mass parameter $m$ we
would get (the $m=0$ branch is listed first)
\begin{align}
\mcm &= (1,0,0,1) \to \left\{\big[0,(1,2),(1,0)\big],\big[1,(3,0),(0,0)\big]\right\}
  & \text{(using the action)} \\
\mcm &= (1,0,0,1) \to \left\{\big[0,(1,2),(1,2)\big],\big[1,(3,0),(1,2)\big]\right\}
  & \text{(using Euler-Lagrange)}
\end{align}

For large $N$ there are more and more ways of splitting the real
fields into specific spin representations; in 4d for $N$ fields
there are $\half(M+1)(M+2)$ different splits, where
$M=\floor(N/4)$.  This grows only mildly as $N^2/32$.  For example,
with $N=5$ there are three splittings: either five spin-0 fields, one
spin-0 and one spin-$\half$, or one spin-0 and one spin-1.  If
we add up all the possible $\mcm$'s for a maximum of $N=20$ real
fields we obtain 160 models, where each model is well defined and
contains all possible interactions that respect the given Lorentz
structure of the kinetic terms.  $N=20$ may be manageable on a
computer, and is just enough to include $\Gsutwo\times\Guone$
gauge theory with a complex Higgs doublet.

The catalogue does not need to be restricted to 4d with
renormalisable couplings.  It really is only limited by one's
imagination and available processing power.  The LPS method can
handle extra dimensions, non-standard kinetic terms, higher-order
operators, supersymmetry, and anything else that can be written
down in an action or with equations of motion.

Although probably not feasible in the near future, we would
ultimately like to construct a catalogue that includes all 4d
theories up to the standard model and beyond, requiring $N$ in the
hundreds.  If the theory beyond the standard model can be
described by an action in 4d then the catalogue would contain this
theory.  From this point of view the putative new theory is part
of a \emph{finite} set (for large but finite $N$).  Using the LPS
method, this is a finite set that we know how to compute.  If
Nature has chosen something highly symmetric then one has the
chance of finding the theory beyond the standard model by sorting
the catalogue on the rank of the symmetry of the branch of each
model, and looking at those with large rank.  Although these ideas
are highly speculative, they provide an alternative perspective on
constructing physics models beyond the standard model.

What we have described in this section is our new approach to
model building.  We are no longer thinking in terms of unification
of gauge groups, what larger groups contain the standard model,
or what matter representations we should choose.  We forget all
that.  \emph{We take the na\"ive perspective of an action as a
bunch of real fields with derivatives, couplings and parameters.
We systematically break down a model into these rudimentary
components, and then use the LPS method to build the model back
up in a systematic way, finding all possible re-constructions, and
all interesting parameter relationships.}


\section{The standard model}
\label{sec:sm}

Perhaps the most obvious thing to do first with the LPS method is
to find all the continuous symmetries of the standard model.
Nature has allowed us to discern this model and we should be
absolutely certain that we are not missing anything.  It may be
that there is something subtle in the standard model that is not
obvious without a systematic exploration.  But it is most likely
that a search reveals only what we already know.  Even so, we
would have then proven the following: as is, the standard
model has no new symmetries and the parameters are pure inputs
with no meaningful relationships.  In order to simplify the
standard model we must extend it.

Consider, in a schematic way, the general structure of the
standard model
\begin{equation}
  \mcl_\text{SM} \sim
    (\partial\phi)^2
    + \phi^2 \partial\phi
    + \phi^2
    + \phi^4
    + \psi \partial\psi
    + \phi \psi^2 \:,
\label{eq:sm-schematic}
\end{equation}
where $\phi$ is a real field with mass dimension $M^1$ and can be
a scalar or gauge field component, and $\psi$ is also a real
field but with dimension $M^{3/2}$ so represents a fermionic
component.  In comparison with~\eqref{eq:nsc-lag}, the second and
fifth terms here are new.  These terms lead to much more involved
structure in the determining equations, an allow for non-Abelian
gauge symmetries and spin-$\half$ representations of the Lorentz
group.  It would be interesting to see if such terms also allow
for new non-compact symmetries relating coordinates and fields.

Written out as a bunch of real fields in the form of
equation~\eqref{eq:sm-schematic}, the standard model with
right-handed neutrinos has $N=244$ real degrees of
freedom,\footnote{We have:
gauge = 4 real components $\times$ (1 hyp + 3 weak + 8 strong) = 48,
leptons = 8 real components $\times$ 3 gens $\times$ ($\nu$ + e) = 48,
quarks = 8 real components $\times$ 3 gens $\times$ 3 cols $\times$ (u + d) = 144,
and Higgs = 2 real components $\times$ weak-doublet = 4.
Total = 244.}
making it a formidable beast indeed.  Using the action approach,
the number of terms in the master determining
equation~\eqref{eq:actde} goes like
$N\times\text{(terms in $\mcl$)} + N^2$.  With approximately $N^2$
terms in $\mcl$, this gives of the order of $10^7$ total terms for
$N=244$.  In this case, the maximum number of determining
equations (set by the number of independent derivatives of fields
with up to three factors, like
$\partial\phi_1\partial\phi_2\partial\phi_3$) is $2.5\times10^6$.
In contrast, starting with the Euler-Lagrange equations
gives orders of magnitude more complexity.  The twice prolonged
operator $\pr^{(2)}\symop$ alone has about $10^9$ terms in it for
$N=244$, and the operator must be applied to 244 equations.
The action approach seems favourable, and there are further
simplifications and tricks we can apply to make it more
manageable.

It would be desirable to generalise the analysis of
Section~\ref{sec:nsc} by adding to the
Lagrangian~\eqref{eq:nsc-lag} the extra terms necessary to
encompass all the terms in the standard model.  In this way it
may be possible to show precisely the symmetries allowed
by the generic Lagrangian~\eqref{eq:sm-schematic}, and further
reduce the determining equations to algebraic form.  This then
makes the final simplification stage pure Gaussian elimination, as
discussed in Section~\ref{sec:auto}.

The number of fields can be reduced by turning off certain parts
of the standard model.  Eliminating the colour and quark sector
leaves only $N=68$ real fields.  Here one could study electroweak
symmetry breaking, lepton family symmetries and neutrino masses
and mixing.  A single generation of only the leptons gives $N=36$,
and a single generation with the colour sector included has
$N=116$.

Unfortunately, we can not use our knowledge of the known
symmetries of the standard model to simplify the analysis.
Additional symmetries mean a more general form of $\eta^\mu$
and/or $\chi_i$, so if one makes an ansatz for these functions
based on known symmetries, then one has immediately excluded the
possibility of finding anything new.

Reducing the determining equations to standard form and obtaining
a set of branches is the most interesting part, and also the most
difficult since the number of branches may become unmanageably
large.  We can decrease the number of branches that are taken by
using our knowledge of the values of the parameters in the
standard model.  At each branch point the numerical value of the
coefficient, which is a function of the parameters, is checked for
zero within the uncertainty of the experimental value of the
parameters.  Only if it is zero within the range is a branch
taken.  If this is manageable, then it should be possible to leave
a couple of parameters completely free and always branch when a
coefficient is dependent on them.  This technique will also be
useful for adding new degrees of freedom to the standard model
with unknown couplings.  One could also linearise all parameters
around their known value, which is easier than the non-linearised
case because one only needs to solve a linear equation at each
branch point.

When checking at a branch point if a combination of parameters
is zero or not, one needs to take into account the fact that
actual parameter values run and depend on the energy scale.
This is not difficult, but needs to be considered.  For example,
one could run the energy scale upward in small steps, and at each
step look for new symmetries, for new branchings that occur due to
the coefficient of a leading term being zero at a particular
energy.  The existence of a new symmetry of the standard model
manifesting at some particular energy scale due to the parameters
unifying is exactly what we expect to happen with unification of
the gauge groups.

The Higgs has not been measured, so to be pedantic one would first
look for the symmetries of the standard model without the Higgs
sector.  Starting with the theory in the electroweak broken phase
with bare mass terms, an LPS search would yield at least Lorentz,
and $\Gsuthree$ and $\Guone$ gauge symmetries.  Moving on from
there, new real degrees of freedom would be added, along with all
their allowed interactions, and the LPS search repeated to look
for additional symmetries.  The important thing to realise is that
adding four real degrees of freedom will reduce the number of
parameters by one and introduce a new gauge symmetry, $\Gsutwo$
(the Higgs mechanism).  Since the LPS method can handle symmetries
written in the broken phase, it should be able to re-discover the
Higgs mechanism in this way.  Going from 240 to 244 degrees of
freedom reduces the parameters by one and increases the total rank
of the symmetries.  An interesting question is whether there
exists a parameter relationship which is not the standard Higgs
mechanism.  The LPS method can give a definitive answer to this,
at least in the regime of adding only a small number of new real
fields.

Assume we have shown this, that we can systematically find the
correct degrees of freedom and interactions that allows an
increase in the symmetry and a reduction in the number of
parameters.  Then we can ask the following question: is it
possible to make the standard model more predictive (reduce the
number of parameters) within the framework of a 4d Lagrangian with
operators?  We can start to answer this question by doing a search
of the standard model plus new degrees of freedom.  If the LPS
method can find the Higgs mechanism, then it should be able to
find the next symmetry group beyond that and the associated
parameter relationship(s).  If, for a large number of new degrees
of freedom, the LPS search comes up empty handed then we conclude
that the standard model cannot be made more predictive using real
degrees of freedom in 4d.  This would point to, for example, the
necessity of extra dimensions and/or supersymmetry, both of which
can be analysed by a more extensive LPS search.  In fact, any
model which can be written as an action or a set of equations of
motion is amenable to the LPS technique of systematically
finding symmetries and relationships among parameters.  The only
limitation is computing power and interpreting the output. 


\section{Conclusions}
\label{sec:concl}

We have described a method for systematically and exhaustively
searching for all continuous symmetries of a model.  The model of
interest can be described by an action, Lagrangian density,
equations of motion, or any set of coupled PDEs.  Using the action
approach, the master determining equation~\eqref{eq:actde}
provides a counterpart to the Euler-Lagrange equations.   It
essentially extracts a linearised version of a theory whose
solutions are the symmetries.  The LPS method also
provides a systematic way to find the solutions of these
equations, or at least their rank.  Along with a list of all
symmetries, the method will also give a systematic list of all
interesting relationships between free parameters (or even free
functions), where interesting means that a different set of
symmetries can be obtained.

The LPS method itself is not new.  What is new is the application
to model building.  Of great interest is to use the method to find
all the symmetries of the standard model, and then go beyond to
find the simplest extensions which yield a reduction in the number
of parameters.  One could also construct a comprehensive catalogue
of all possible 4d field theories for low numbers of real degrees
of freedom.

In the 35 or so years since the standard model was written down,
there has not been one model which is more predictive and reduces
the number of parameters (disregarding the neutrino sector).  But
there have been countless attempts at this.  Attempts to
intuitively guess a bigger symmetry group and then from it derive
the standard model and its couplings.  The method described in
this paper gives a new approach to model building by providing a
much more systematic way to search for extensions of, for example,
the standard model.  Given a particular extension of degrees of
freedom and couplings, the method allows one to find all possible
symmetries and \emph{derive} constraints among parameters,
removing part of the guess work in model building.


\begin{acknowledgments}
I would like to thank M.~Postma, B.~Schellekens, R.~de Adelhart
Toorop and S.~Mooij for stimulating discussions and reading a past
manuscript.  I would also like to thank R.R.~Volkas, R.~Davies and
K.L.~McDonald for very helpful comments on a previous version of
this paper.  This research was supported by the Netherlands
Foundation for Fundamental Research of Matter (FOM) and the
Netherlands Organisation for Scientific Research (NWO).
\end{acknowledgments}



\providecommand{\href}[2]{#2}\begingroup\raggedright\endgroup

\end{document}